\pdfoutput=1
\documentclass[iop,revtex4]{emulateapj}
\usepackage{graphicx,subfigure}
\usepackage{float}
\usepackage{color}
\usepackage{amsmath}
\usepackage[english]{babel}
\usepackage{epstopdf}
\usepackage{amsmath,amsfonts,xspace,textcomp,graphicx}

\slugcomment{Submitted to the Proceedings of the Astronomical Society of the Pacific}

\newcommand{\muas}{\mbox{{\usefont{U}{eur}{m}{n}{\char22}}as}\xspace}

\shorttitle{Residual atmospheric dispersion measurement and correction}
\shortauthors{P. Pathak et al.}

\begin{document}
\title{A high precision technique to correct for residual atmospheric \\ dispersion in high-contrast imaging systems}
\author{P. Pathak\altaffilmark{1}, O. Guyon\altaffilmark{1,3,4}, N. Jovanovic\altaffilmark{1,6}, J. Lozi\altaffilmark{1}, F. Martinache\altaffilmark{2}, Y. Minowa\altaffilmark{1}, \\T. Kudo\altaffilmark{1}, H. Takami\altaffilmark{5}, Y. Hayano\altaffilmark{5}, N. Narita\altaffilmark{5,7,8} }

\altaffiltext{1}{National Astronomical Observatory of Japan, Subaru Telescope, 650 North A'Ohoku Place, Hilo, HI, 96720, U.S.A.}
\altaffiltext{2}{Observatoire de la Cote d'Azur, Boulevard de l'Observatoire, Nice, 06304, France}
\altaffiltext{3}{Steward Observatory, University of Arizona, Tucson, AZ, 85721, U.S.A.}
\altaffiltext{4}{College of Optical Sciences, University of Arizona, Tucson, AZ 85721, USA}
\altaffiltext{5}{National Astronomical Observatory of Japan, 2-21-1 Osawa, Mitaka, Japan}
\altaffiltext{6}{Department of Physics and Astronomy, Macquarie University, Sydney, NSW 2109, Australia}
\altaffiltext{7}{Department of Astronomy, The University of Tokyo, 7-3-1 Hongo, Bunkyo-ku, Tokyo, 113-0033, Japan}
\altaffiltext{8}{Astrobiology Center, 2-21-1 Osawa, Mitaka, Tokyo, 181-8588, Japan}
\email{ppathak8@gmail.com}


\begin{abstract}
Direct detection and spectroscopy of exoplanets requires high contrast imaging. For habitable exoplanets in particular, located at small angular separation from the host star, it is crucial to employ small inner working angle (IWA) coronagraphs that efficiently suppress starlight. These coronagraphs, in turn, require careful control of the wavefront which directly impacts their performance. For ground-based telescopes, atmospheric refraction is also an important factor, since it results in a smearing of the PSF, that can no longer be efficiently suppressed by the coronagraph.  Traditionally, atmospheric refraction is compensated for by an atmospheric dispersion compensator (ADC). ADC control relies on an a priori model of the atmosphere whose parameters are solely based on the pointing of the telescope, which can result in imperfect compensation. For a high contrast instrument like the Subaru Coronagraphic Extreme Adaptive Optics (SCExAO) system, which employs very small IWA coronagraphs, refraction-induced smearing of the PSF has to be less than 1 mas in the science band for optimum performance. In this paper, we present the first on-sky measurement and correction of residual atmospheric dispersion. Atmospheric dispersion is measured from the science image directly, using an adaptive grid of artificially introduced speckles as a diagnostic to feedback to the telescope's ADC. With our current setup, we were able to reduce the initial residual atmospheric dispersion from 18.8 mas to 4.2 in broadband light (y- to H-band), and to 1.4 mas in H-band only. This work is particularly relevant to the upcoming extremely large telescopes (ELTs) that will require fine control of their ADC to reach their full high contrast imaging potential.
\end{abstract}

\section{Introduction}\label{s:intro}
Several thousand exoplanets have thus far been discovered using indirect methods, such as transit and radial velocity but very few using direct imaging\footnotemark \footnotetext{exoplanets.org}. To answer questions about habitability of exoplanets, it is essential to utilize direct detection methods in order to be able to conduct spectroscopic studies. With advances in technology, direct detection of exoplanets is becoming more robust and mature. High contrast instruments like the Gemini Planet Imager (GPI) \citep{macintosh14} and the Spectro-Polarimetric High-contrast Exoplanet REsearch instrument (SPHERE) \citep{bez08}, are now able to perform surveys, and have already detected new planets \cite{bruce15,wagner2016}. 

The basic architecture of a high contrast instrument employs extreme adaptive optics (ExAO) to compensate for atmospheric turbulence and coronagraphs to suppress the light coming from the host star. A potentially significant source of coronagraphic leakage comes from atmospheric refraction once the turbulence has been accounted for, which results in an elongated point spread function (PSF) that limits the achievable angular resolution and coronagraphic contrast. This effect can be particularly devastating if the coronagraph operates at a small inner working angle (IWA). 

The PSF elongation for previous generation high contrast instruments such as the High Contrast Instrument for the Subaru Next Generation Adaptive Optics (HiCIAO)~\citep{hiciao} was specified to be $1/30^{th}$ of the diffraction-limit ($1\lambda/D=40$~mas in H-band, residual dispersion goal was 1.33~mas): our experiments indicate that this was not achieved in practice (the measured residual dispersion was $\sim12$~mas in H-band). The latest high contrast instrument, the Subaru Coronagraphic Extreme Adaptive Optics (SCExAO) system \citep{jovanovic15}, which is constantly evolving with time by employing the latest technologies, is going a step further in the field of direct imaging. The science requirement for SCExAO is a spread in the PSF of $\sim1/50^{th}$ of the diffraction-limit, which translates to $<1$~mas in H-band. At this level the system will no longer be limited by residual atmospheric dispersion, instead the finite stellar angular size will become the dominant term contributing to stellar leakage. For example, a solar-mass star at 10~pc has an angular size of 1~mas, demonstrating that stars normally considered as unresolved are actually partially resolved and will affect performance at these levels. The correction of the residual atmospheric dispersion is done using an atmospheric dispersion compensator (ADC), which consists of two prisms with similar dispersive properties. The amount and direction of the compensation are adjusted by rotating the prisms with respect to one another and rotating the entire assembly around the axis of propagation \citep{allen}.

An ADC model basically uses the local temperature and pressure to determine the index of refraction of air around the telescope. It does this at multiple wavelengths to guess the dispersion for different angles at the telescope. To get a correction of 1 mas ($10^{-3}$~arcsec), since the atmospehric dispersion accross visible and IR is about 1~arcsec, one needs to measure the local temperature and pressure to at least a $10^{-3}$ precision. In practice, this precision is never achieved. So although the model include these parameters precisely, the inputs may not achieve the right precision. One other term that complicates things is the humidity. Indeed, water vapor changes the index of the air. The quantity of water vapor is a key factor, as well as the wavelength observed. This effect cannot be taken into account easily. The problem with the current approach of correction is that it does not make any measurement of the actual dispersion and hence results in imperfect correction. 

Adaptive optics (AO) systems include model based ADCs, such as Keck AO \citep{wynne} and Subaru's AO188 \citep{subaruadc}. In this paper, we present the first on-sky measurement and correction of residual atmospheric dispersion. Here we demonstrate that by using a grid of speckles generated by the deformable mirror (DM) of the SCExAO instrument, we can accurately measure the residual atmospheric dispersion and subsequently correct it to $\approx 1.4 $~mas in H-band. In Sec.~2 and 3, we show the concept and its implementation using simulation, and Sec.~4 presents the on-sky results.  

\section{Principle}\label{s:principle}
\subsection{Measuring Atmospheric Dispersion}
\begin{figure}
  \centerline{
    \resizebox{0.47\textwidth}{!}{\includegraphics{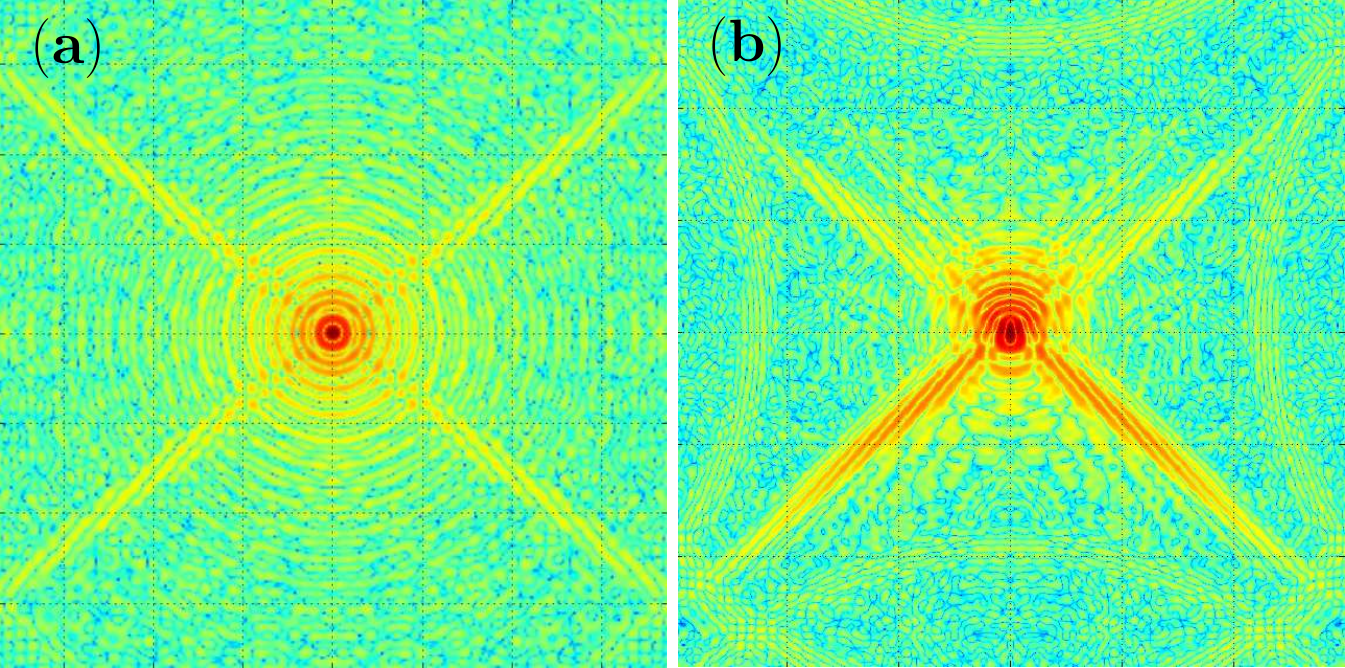}}}
  \caption{Simulated PSF with bandwidth covering y to H-band: (a) With no atmospheric dispersion, i.e. telescope pointing at zenith. (b) With atmospheric dispersion corresponding to a telescope elevation of $35^\circ$.}
  \label{f:psf_0}
\end{figure}

Light from an astronomical object initially undergoes refraction at the space/atmosphere boundary and then continues to refract as it propagates towards the ground. This continuous refraction is due to the increasing density and temperature of the atmosphere as one moves to lower altitudes and results in an arc-like trajectory for the light rays as they traverse the atmosphere.  Refraction of course, is a function of wavelength and hence a polychromatic PSF will look elongated at the focus of a telescope if it is not pointing at zenith. To illustrate this point, Fig.~\ref{f:psf_0} shows an example of a simulated PSF (y- to H-band) for a telescope pointing at zenith with no atmospheric dispersion and compares it to an image using the same conditions, only at a lower elevation of $35^\circ$, which results in strong atmospheric dispersion. Over a broad bandwidth ($>10$~nm or so), atmospheric refraction results in an elongation in the PSF, as shown in the Figure~\ref{f:psf_0} (b). An ADC, relying on a model of the atmosphere effectively compensates for most of this refraction, reducing the elongation of this PSF. However, without a direct diagnostic tool, some amount of residual dispersion remains. One could attempt to diagnose the residual dispersion (20-100~\muas/nm) by directly measuring the PSF elongation, but this measurement is difficult in practice at the level required by the coronagraph due to residual wavefront error ($<1$~mas over the H-band). Instead of attempting to measure dispersion in the PSF core itself, here we suggest using the properties of diffraction features at large separation ($>10$~$\lambda/D$) whose location is a more sensitive function of the refraction.

\begin{figure}
  \centerline{
		\resizebox{0.47\textwidth}{!}{\includegraphics{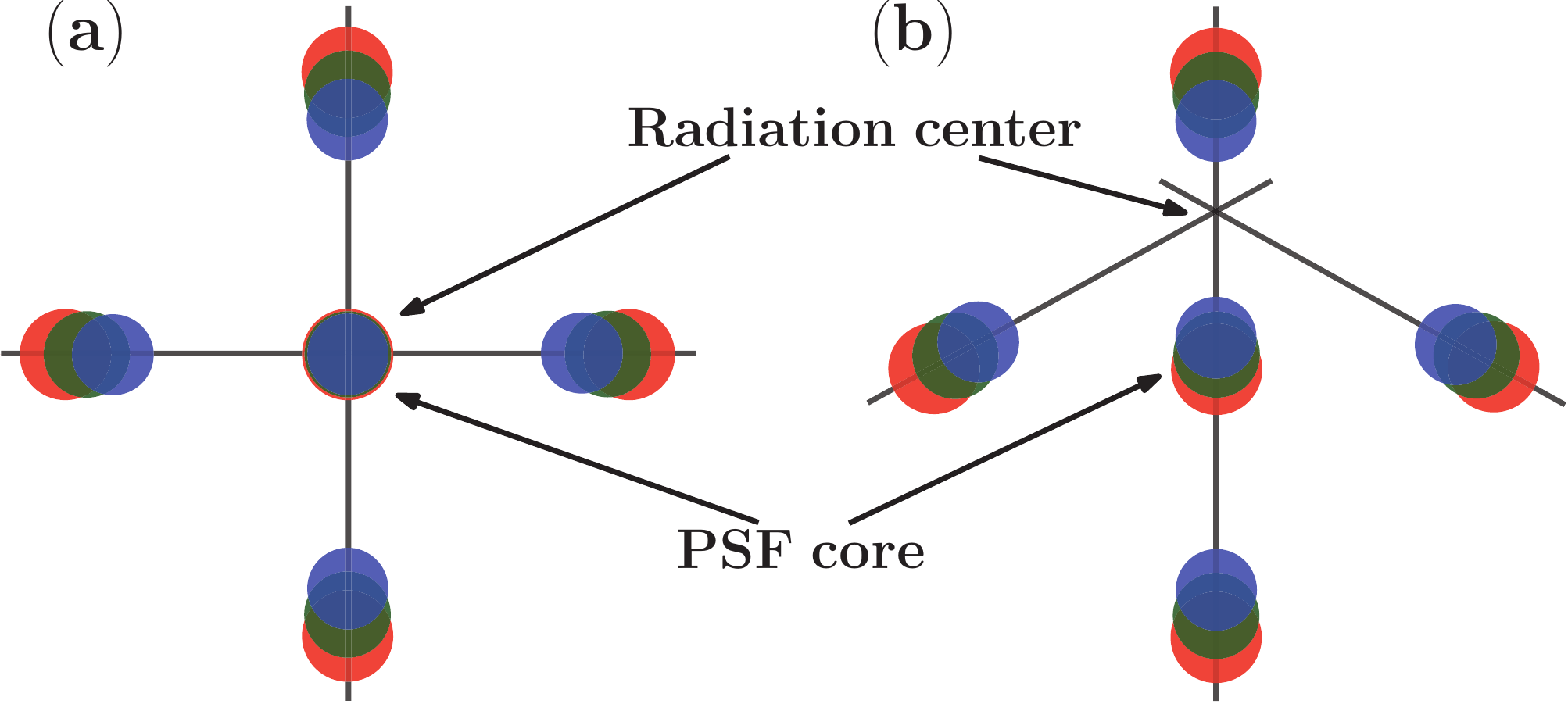}}}
	\caption{Concept of the measurement of atmospheric dispersion:
		The on-axis PSF is diffracted to generate speckles. (a) No dispersion, each wavelength is focused at the same point on-axis, and the lines joining the diffracted speckles meet at the PSF core. (b) With atmospheric dispersion, the different wavelengths are no longer centered, and the lines joining the diffracted speckles meet at radiation center.}
	\label{f:schematic}
\end{figure}

Figure~\ref{f:schematic} presents the rationale behind our approach. It features a schematic representation of a broad-band PSF with conveniently placed symmetric off-axis diffraction features (speckles), and shows how these are affected by the presence of refraction. With no refraction (corresponding to pointing at zenith), the images at all wavelengths will radiate from a point that coincides with the PSF core (cf. Fig.~\ref{f:schematic} left). Unlike the core of the PSF, off-axis diffraction features appear spectrally dispersed, with the shortest wavelengths closer to the center of the image (cf. Fig.~\ref{f:schematic} (a)). When affected by residual differential refraction, the images at different wavelengths no longer radiate from a point coinciding with the PSF core (cf. Fig.~\ref{f:schematic} (b)), but from a point that we will call the radiation center, offset from the PSF core. By measuring the distance between the radiation center and the PSF core, one can directly estimate the amount of residual dispersion. 

\subsection{Correcting Atmospheric Dispersion}
\begin{figure}
	\centerline{
	\resizebox{0.48\textwidth}{!}{\includegraphics{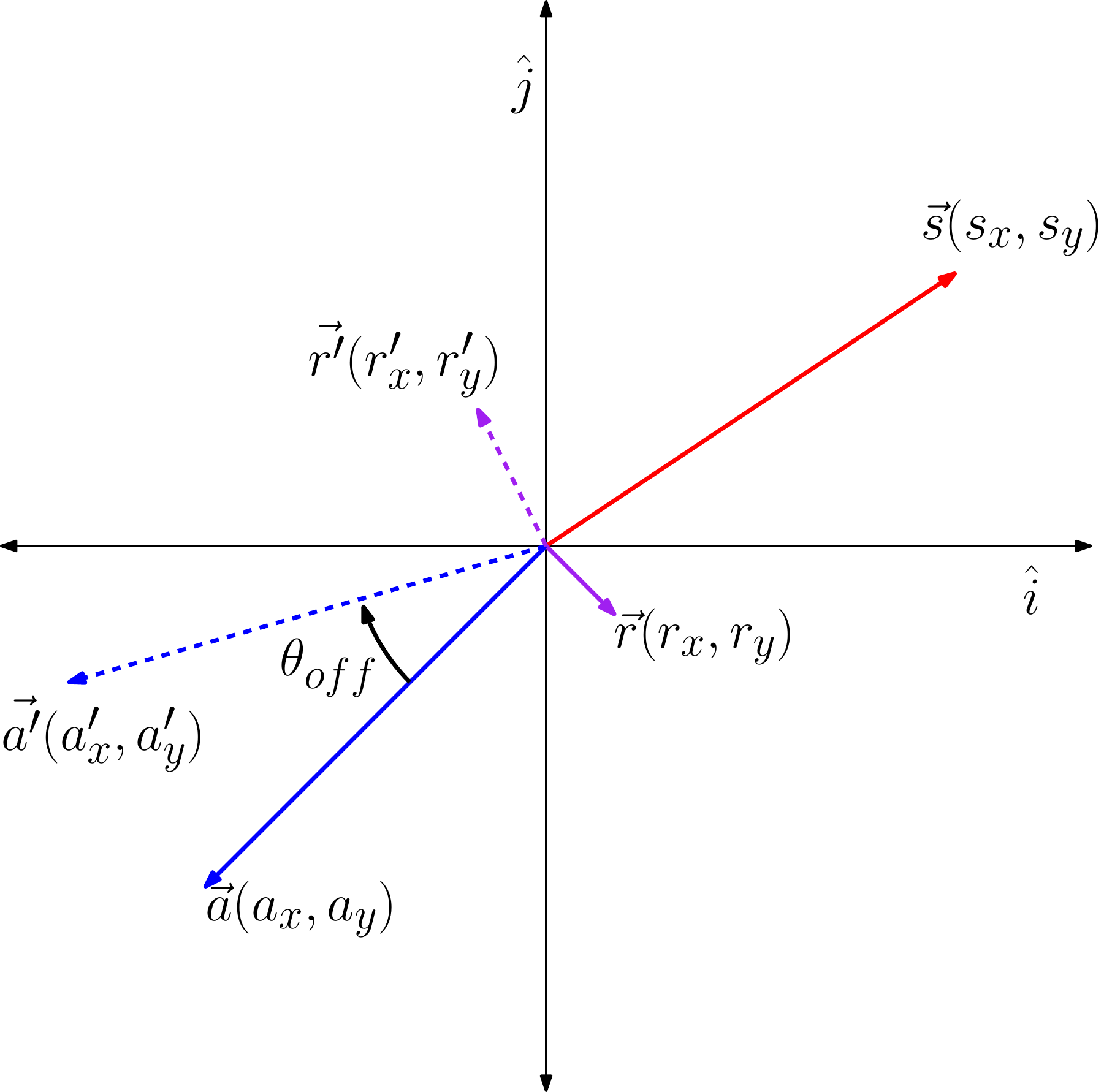}}}
	\caption{Principle of the ADC: the dispersion vector $\vec{\bf s}$ is partially canceled by the ADC vector $\vec{\bf a}$ and the residual vector is given by $\vec{\bf r}$. ADC vector $ \vec{\bf a'}$ is generated when $\vec{\bf a}$ is offset by an angle $\theta_{off}$, resulting in a new residual vector $ \vec{\bf r'}$.}
	\label{f:vector}
\end{figure}

Figure~\ref{f:vector} shows a vector based representation of the effect of atmospheric dispersion. The axes correspond to cartesian coordinates in the image plane. The atmospheric dispersion vector, i.e. the direction in which the PSF is elongated, is designated with the symbol $\vec{\bf s}$ in the figure. The dispersion vectors for the two prisms of the ADC are represented by $ \overrightarrow{\bf p_1} $ and $ \overrightarrow{\bf p_2}$. They have equal dispersion magnitudes and add together to generate the total ADC vector $\vec{\bf a}$, which is also dependent on orientation and is given by

\begin{align}
\vec{\bf a} &= \overrightarrow{\bf p_1} + \overrightarrow{\bf p_2} \label{eqn1}.
\end{align}

If $p$ is the dispersion magnitude of each prism, and $\theta_1 $ and $\theta_2$ their orientation angles, then the two individual prism dispersion vectors are decomposed in the image coordinate system as

\begin{equation} \label{eq2}
\left\{
\begin{array}{rl}  
\overrightarrow{\bf p_1} &= p \cos(\theta_1) \hat{i} + p \sin(\theta_1) \hat{j} \\
\overrightarrow{\bf p_2} &= p \cos(\theta_2) \hat{i} + p \sin(\theta_2) \hat{j}
\end{array}
\right..
\end{equation}

The ADC provides a maximum dispersion when the two prisms are aligned. When one prism is rotated 180 degrees relative to the other (anti-aligned), the ADC results in zero compensation. For more details of how an ADC operates, refer to~\citet{subaruadc}.

In the case of an incomplete compensation, what is left is the residual atmospheric dispersion vector $\vec{\bf r}$, as shown in Fig.~\ref{f:vector}, given by

\begin{align}
\vec{r} = \vec{a} + \vec{s} \label{r}.
\end{align}

The goal is to measure the residual dispersion and offset the ADC to minimize the dispersion, in a closed-loop operation. The steps involve the calibration of the response of the ADC prisms (prism dispersion magnitude $p$) and the calculation of on-sky dispersion. If we assume that the ADC prism angles are known all the time, then the components in the image coordinate system can be written as

\begin{equation} \label{ax}
\left\{
\begin{array}{rl}  
a_x &= p \cos(\theta_1) + p \cos(\theta_2)\\
a_y &= p \sin(\theta_1) + p \sin(\theta_2)
\end{array}
\right..
\end{equation}

In the same coordinate system, the residual vector can be decomposed into

\begin{equation} \label{1}
\left\{
\begin{array}{rl}  
r_x &= s_x + a_x \\
r_y &= s_y + a_y . 
\end{array}
\right..
\end{equation}

To calibrate the response of the ADC prisms, we assume that, over small timescales, the on-sky dispersion vector $\vec{\bf s}$ is constant. As shown in Fig.~\ref{f:vector}, by making two measurements of residual $\vec{\bf r}$ and $\vec{\bf r'}$ using two known positions of ADC, we can eliminate the atmospheric contribution $\vec{\bf s}$. By rotating the ADC by an angle $\theta_{off}$, the new ADC vector $\vec{\bf a'}$ is given by
\begin{equation} \label{a'x}
\left\{
\begin{array}{rl}  
a'_x &= p \cos(\theta_1-\theta_{off}) + p \cos(\theta_2-\theta_{off})\\
a'_y &= p \sin(\theta_1-\theta_{off}) + p \sin(\theta_2-\theta_{off})
\end{array}
\right..
\end{equation}

The new residual vector $\vec{\bf r'}$ is then decomposed into
\begin{equation} \label{2}
\left\{
\begin{array}{rl}   
r'_x &= s_x + a'_x  \\
r'_y &= s_y + a'_y  
\end{array}
\right..
\end{equation}

By subtracting Eq.~\ref{2} from Eq.~\ref{1}, we get
\begin{equation}\label{11}
\left\{
\begin{array}{rl}
r_x - r'_x &= a_x - a'_x  \\
r_y - r'_y &= a_y - a'_y 
\end{array}
\right..
\end{equation}

Now substituting the Eqs.~\ref{ax} and \ref{a'x} into Eq.~\ref{11}, we get
\begin{equation}\label{5}
\left\{
\begin{array}{rl}
p \times l &= r_x - r'_x \\
p \times m &= r_y - r'_y 
\end{array}
\right.,
\end{equation}
where $l$ and $m$ are given by
\begin{equation} \label{lm}
\left\{
\begin{array}{rl}  
l &= \cos(\theta_1)+\cos(\theta_2)-\cos(\theta_1-\theta_{off})\\
& \ \ \ -\cos(\theta_2-\theta_{off})\\
m &= \sin(\theta_1)+\sin(\theta_2)-\sin(\theta_1-\theta_{off})\\
& \ \ \ -\sin(\theta_2-\theta_{off})
\end{array}
\right..
\end{equation}

By solving Eq.~\ref{5}, we get 
\begin{align}
p = \left(\frac{(r_x - r'_x)^2+(r_y - r'_y)^2}{l^2+m^2}\right)^{\frac{1}{2}} \label{9}.
\end{align}

Since we know the prism angles $\theta_1 $ and $ \theta_2$,  the offset applied to the prisms $\theta_{off}$, the value of $l$ and $m$ can be deduced using Eq.~\ref{lm}. Then by substituting the values of the measured residual dispersion $(r_x,r_y) $, $ (r'_x,r'_y)$, and $l$ and $m$ into Eq.~\ref{9}, the magnitude of the prism dispersion vector $p$ can be calculated. 

Once $p$ is determined from the measurements, the vector $\vec{\bf a}$ is known for any ADC angle using Eq.~\ref{r}. For any measurement of the residual vector $\vec{\bf r}$, we can deduce the on-sky dispersion vector $\vec{\bf s}$ using Eq.~\ref{r}. Finally, once we know the dispersion $\vec{\bf s}$, we can determine the new ADC angle that compensates for it.

\begin{figure}
	\centerline{
		\resizebox{0.48\textwidth}{!}{\includegraphics{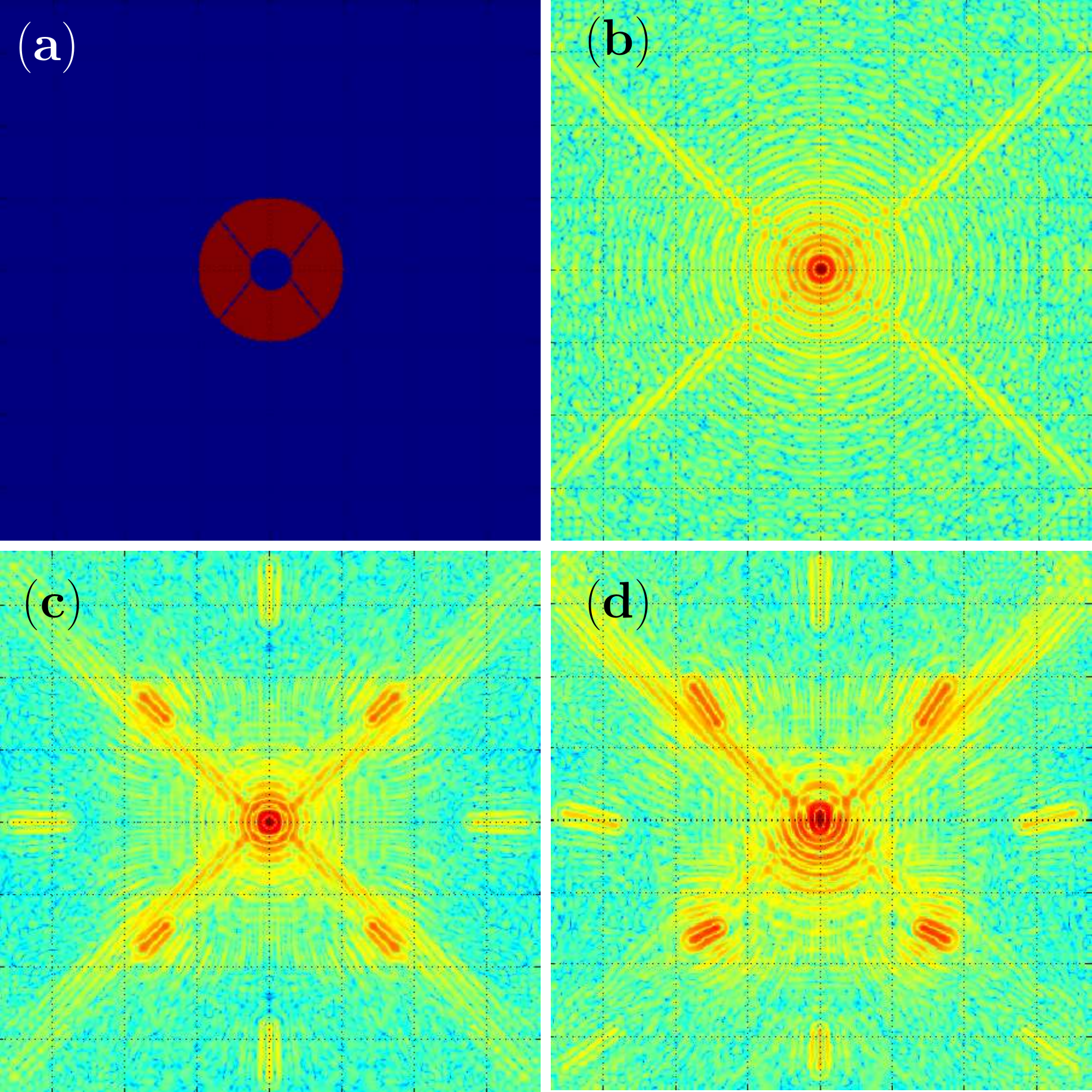}}}
	\caption{The theoretical pupil and corresponding PSFs assuming no turbulence applied (i.e. off-sky). All PSF images are polychromatic. (a) Subaru Telescope pupil geometry. (b) PSF corresponding to the pupil. (c) Same PSF including four artificial speckles generated by applying a sinusoidal modulation in phase across the pupil. (d) Same PSF with speckles including the affect of atmospheric dispersion.}
	\label{f:psf}
\end{figure}

\section{Simulation} \label{simulation}
To test the concepts explained in the previous section, we present the result of simulations that further characterize our approach.  As explained by \citet{jovanovicsp15} and \citet{frantz14}, artificial speckles at specific locations can deliberately be introduced in an image by a DM featuring a sufficiently large number of actuators, within a finite region called the control region of the DM. We use this convenient approach to generate symmetric pairs of speckles in the image that will be used to sense the amount of residual refraction in focal plane images.

The simulations take the Subaru telescope pupil geometry into consideration, as shown in Fig.~\ref{f:psf} (a). For the images in Fig.~\ref{f:psf} (b), (c) and (d), a 12.1~mas/pix plate scale was used to match that of SCExAO's internal NIR camera. Figure~\ref{f:psf} (b), shows what a perfect monochromatic PSF looks like.

Panels (c) and (d) of Fig.~\ref{f:psf} feature simulated broadband images, corresponding to the combined y- to H-band flux. Low-amplitude sinusoidal modulations were added to the DM to produce pairs of off-axis speckles along the diagonals of $x$ and $y$ axes of the image. For panel (c), no differential refraction is implemented, so that the spectrally dispersed off-axis speckles appear to radiate from the core of the PSF. In panel (d), 220~\muas/nm of differential refraction was introduced: the radiation center of the off-axis speckles is no longer pointed toward the PSF core.
\begin{figure}
	\centerline{
		\resizebox{0.48\textwidth}{!}{\includegraphics{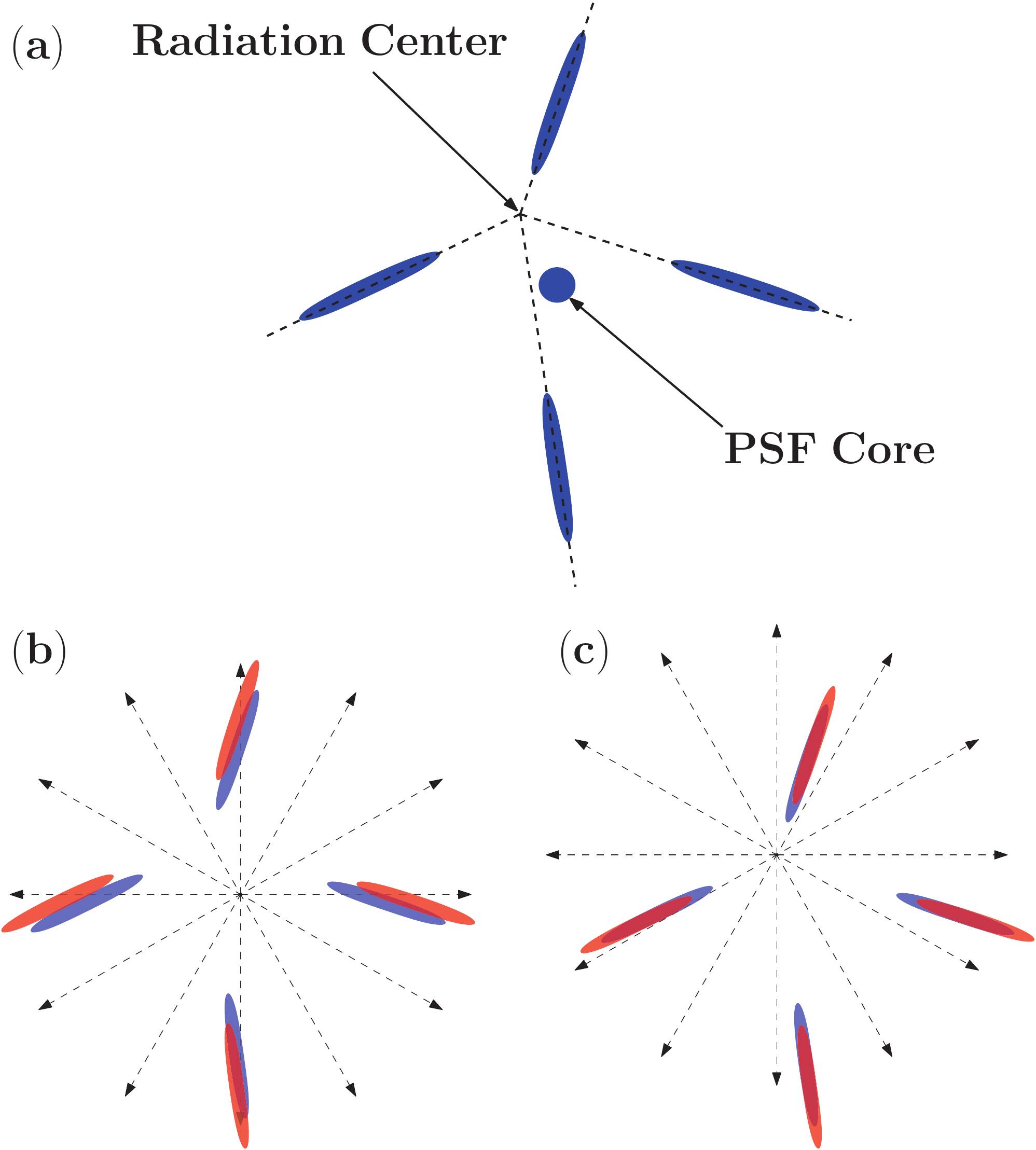}}}
	\caption{(a) PSF core and speckles with the presence of atmospheric dispersion. The elongated speckles meet at the radiation center. (b) Speckles only with PSF core removed. The blue speckles represent the original speckles, while the red ones are the same speckles, stretched radially from the PSF core. (c) Same schematic as in (b), except the stretch is done from the actual radiation center.}
	\label{f:sch1}
\end{figure}
\subsection{Extracting Atmospheric Dispersion} \label{ssec:num1}
In order to estimate the amount of residual atmospheric refraction in a given image, the distance between the core of the PSF and the radiation center must be measured precisely. Even with some prior knowledge, the careful mapping of non-symmetric structures is a non trivial task in practice. We have however empirically developed an algorithm that provides satisfactory performance which is summarized in Fig.~\ref{f:sch1}: 

\begin{itemize}
\item Panel (a) shows a schematic representation of an image affected by residual refraction. We use $\mathrm{d_x}$ and $\mathrm{d_y}$ to refer to the horizontal and vertical offset between the PSF core and the radiation center respectively.
\item Panel (b) shows a similar representation for which the PSF core has been masked out. The initial guess is that the PSF core corresponds to the radiation center (i.e. there is no residual dispersion). The original (blue) image is then radially stretched from the location of the original PSF core by a finite amount (in red). One can observe that for a sufficiently large stretch factor (of the order of 20\% in practice), the spectrally dispersed structures present in both images (the blue and the red) no longer overlap. This difference can be quantified by the norm of the difference between the two images $\mathcal{L}$.
\item The next step is to shift the assumed center of radiation, and re-stretch the original image to find the point at which the norm $\mathcal{L}$ is minimized (i.e. the speckles from the original image and those from the stretched image overlap most). In this case a raster scan is conducted but other scanning algorithms could also be used. 
\item Panel (c) shows a final case featuring both the original (blue) and the stretched (red) version of the images, only this time, the latter was stretched from the correct radiation center: the speckles now have a significant overlap and our criterion of minimizing $\mathcal{L}$ is satisfied.
\end{itemize}

\begin{figure}
	\centerline{
		\resizebox{0.45\textwidth}{!}{\includegraphics{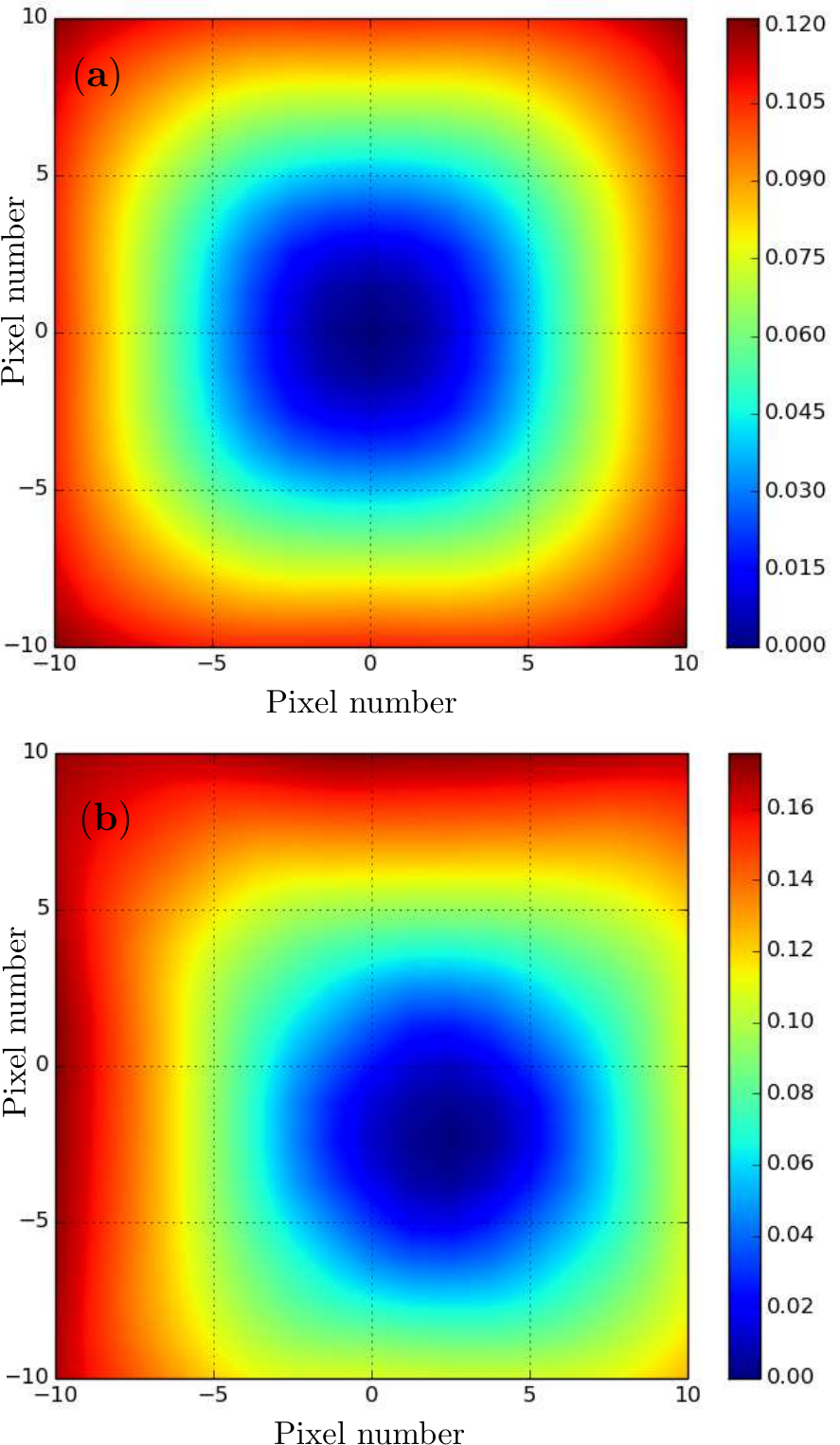}}}
	\caption{Offset of the radiation center from the PSF core, with the color bar representing the relative intensity: (a) with no atmospheric dispersion, (b) with atmospheric dispersion of 18.8~\muas/nm.}
	\label{f:sim1}
\end{figure}

The result of one such computation is shown in Fig.~\ref{f:sim1} for two cases, in the absence of residual refraction in panel (a), and with 18.8~\muas/nm of refraction in panel (b). As expected in the first case, $\mathcal{L}$ is minimum when the PSF core and the radiation center are identical (i.e. $\mathrm{d_x} = 0$ and $\mathrm{d_y} = 0$). In the second scenario, the location of the minimum of the $\mathcal{L}$ is offset. The distance between the PSF core and the radiation center is about twice this offset, i.e.
\begin{align}
r_i \simeq 2 \times d_i \quad i \in [x,y] \label{ratio}.
\end{align}
This relationship is used to calculate the on-sky dispersion in Sec.~\ref{on-sky}. By this method one can measure the residual dispersion down to sub-pixel accuracy.

\begin{figure}
	\centerline{
		\resizebox{0.48\textwidth}{!}{\includegraphics{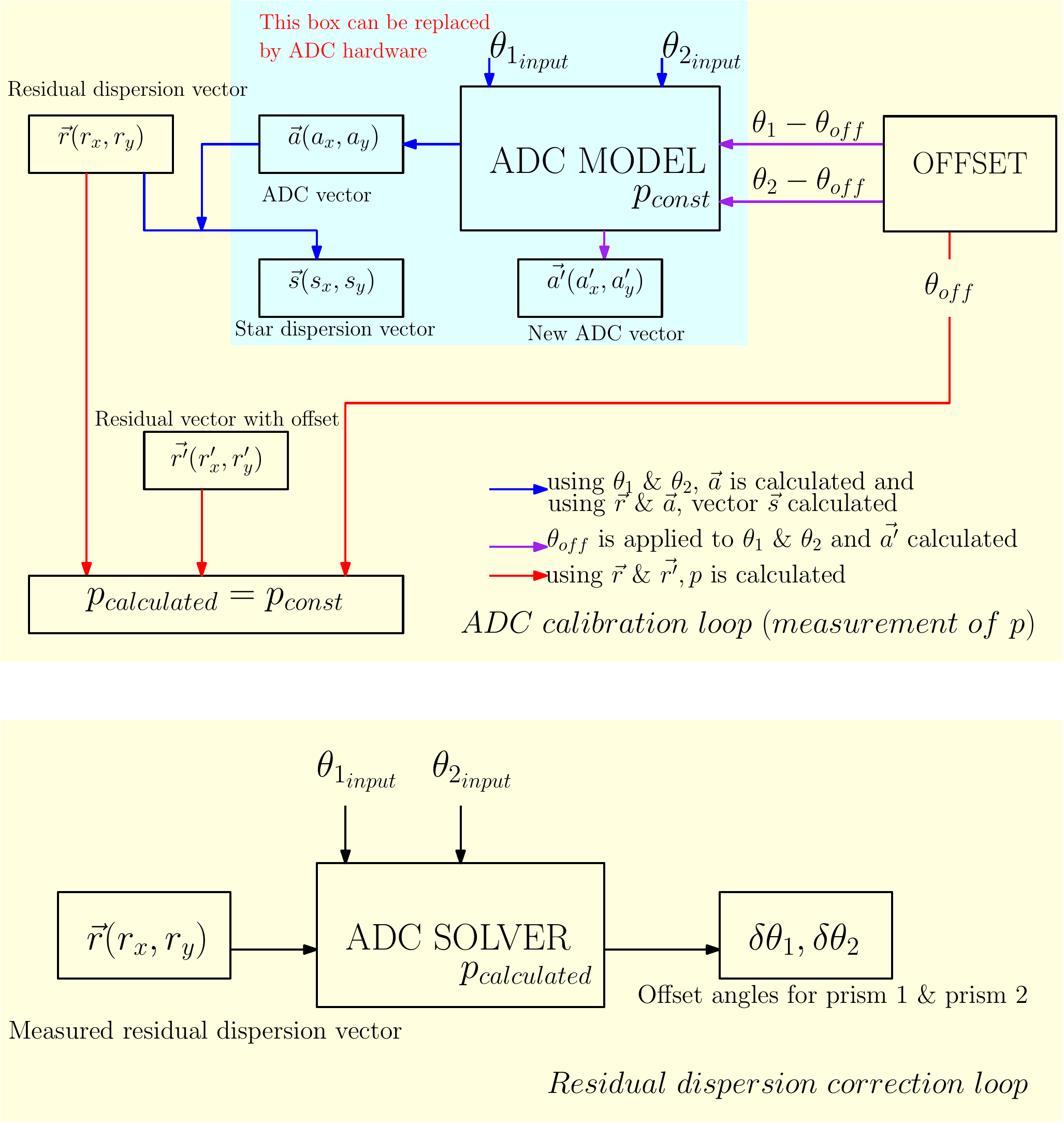}}}
	\caption{Schematic of the control loop, used for calibration of the ADC and calculating the offset angles of the prisms, and for correcting the atmospheric dispersion.}
	\label{f:adc}
\end{figure}

\subsection{ADC Simulation}
The operation of an ADC was simulated to test the measurement and correction of the dispersion on a simulated PSF. As described in Fig.~\ref{f:adc}, the first part of the simulation involves the calibration of the ADC. The inputs used in simulation are prism angles $\theta_1 $ and $\theta_2$ and the magnitude of the prism dispersion vector $p$, which is an unknown constant. By using these parameters, the ADC vector $\vec{\bf a}$ is calculated, using the equations presented in Sec.~\ref{s:principle}. By using the measured vectors $\vec{\bf r} $ and $\vec{\bf a}$, the on-sky dispersion vector $\vec{\bf s}$ can be calculated. In a second step, an offset $\theta_{off}$ is applied to $\theta_1 $ and $\theta_2$, which gives the new ADC vector $\vec{\bf a'}$ and residual vector $\vec{\bf r'}$. By using residual vectors $\vec{\bf r}, \ \vec{\bf r'}$ and the offset angle $\theta_{off}$, the prism dispersion magnitude $p$ can be calculated. In the second part of the simulation using the measured residual dispersion vector $\vec{\bf r}$, the prism angles and $p_{calculated}$, offset angles $ \delta \theta_1 $ and $ \delta \theta_2$ are calculated, that will give us a better correction of dispersion. 
\begin{figure}
	\centerline{
		\resizebox{0.48\textwidth}{!}{\includegraphics{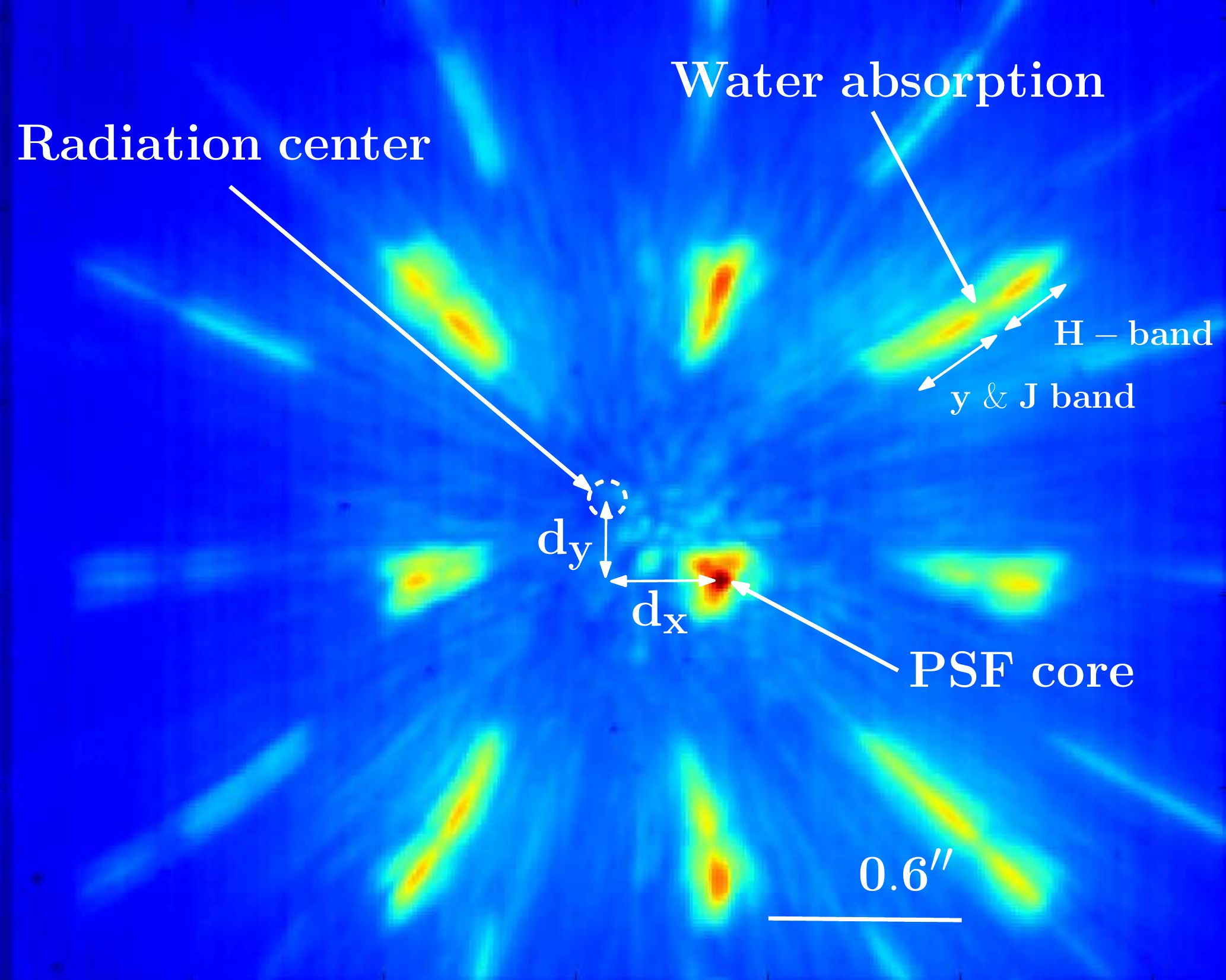}}}
	\caption{On-sky PSF, showing the radiation and PSF core. The deviation of the radiation center from the location of the PSF core shows the presence of atmospheric dispersion. Here, a large offset is deliberately added to show the principle.}
	\label{f:onsky}
\end{figure}

\section{Results} \label{on-sky}
As explained in \citet{jovanovicsp15} and \citet{frantz14}, artificial speckles can be generated by diffracting the PSF core with a DM having a large number of actuators. In this case, SCExAO's DM was used, which has a total of 2000~actuators, with 45~actuators across the pupil. The DM can be modulated to create a grating structure in the form of a sine wave. The distance between the PSF core and the resulting speckles is a function of the number of periods across the pupil. The more cycles per aperture the sine wave has, the further a speckle is projected from the PSF. With 45~actuators across the pupil, the furthest speckles can be placed is 22.5~$\lambda/D$ from the PSF \citep{jovanovic15}. The brightness of the speckles can be controlled by adjusting the amplitude of the sine wave.

Previously, this ability to arbitrarily generate speckles has been used to systematically remove speckles in the PSF Halo~\citep{frantz14} and more recently for high precision astrometry using incoherent speckles~\citep{jovanovicsp15}. For our tests, on-sky speckles were placed at 22.5~$\lambda/D$, with a 100~nm RMS amplitude. Images were taken using a near infrared (NIR) camera ($320\times256$~pix InGaAs detector). The target was Beta Leo (spectral type A3, R-mag$=2.08$, H-mag$=1.92$) on SCExAO's engineering night of April 2$^{nd}$, 2015. The data was collected after AO188, the Subaru Telescope facility AO instrument. In closed-loop, it offers Strehl ratios in H-band between 20 and 40\% \citep{minowa10}. To capture the elongated speckles, the light from y- to H-band was captured by the NIR camera. For image processing, an averaged dark was calculated from a cube of 1000~dark frames, then subtracted from the science images. The hot pixels were also removed. At the time of the test, the telescope elevation was $43^\circ$ and no ADC correction was applied to highlight the presence of atmospheric dispersion. As shown in Fig.~\ref{f:onsky}, the speckles do not point to the PSF core as expected. 
\begin{figure}
	\centerline{
		\resizebox{0.48\textwidth}{!}{\includegraphics{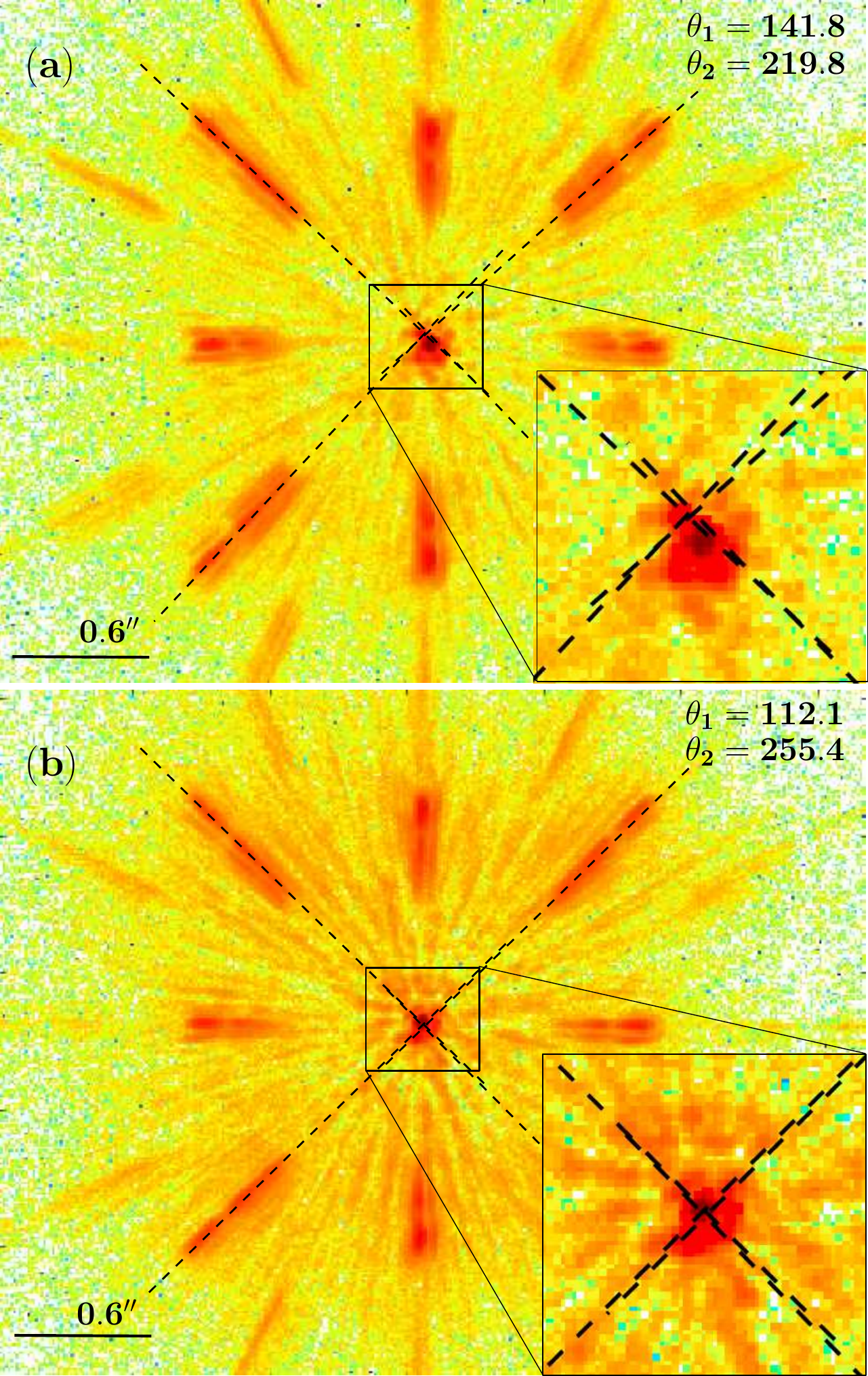}}}
	\caption{PSF with added speckles, $\theta_1$ and $\theta_2$ give the actual prisms angle. (a) Before correcting the residual atmospheric dispersion (fitted lines are over-plotted to show that the radiation center does not coincide with the PSF core). (b) After correcting the residual dispersion (the fitted lines show the radiation center coincides with the PSF core).}
	\label{f:result}
\end{figure}
\begin{figure}
	\centerline{
		\resizebox{0.45\textwidth}{!}{\includegraphics{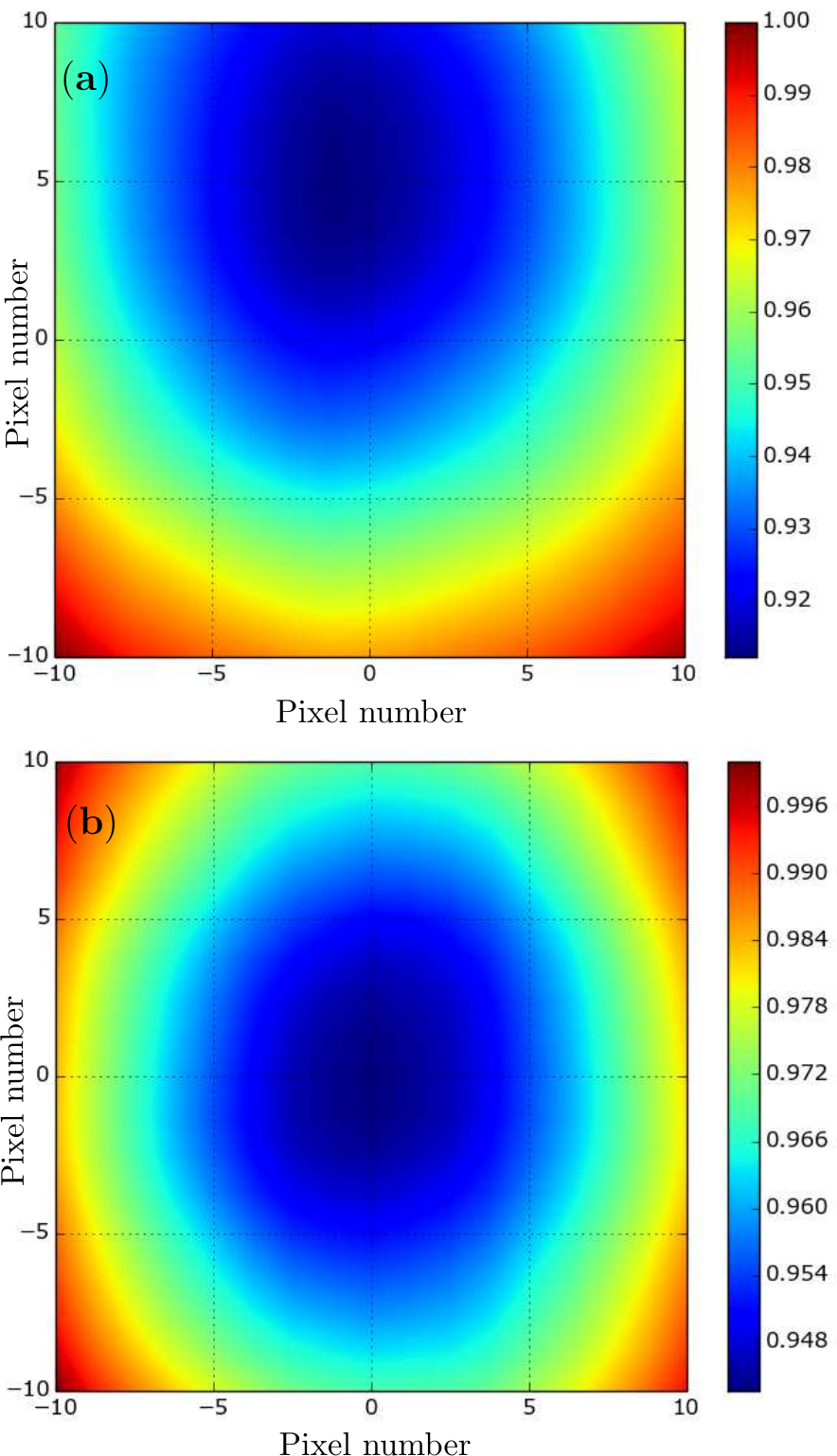}}}
	\caption{On-sky results: Offset of the radiation center from PSF core. The colorbar represents the relative intensity. (a) Before correcting the residual atmospheric dispersion. (b) After correcting the residual atmospheric dispersion.}
	\label{f:result1}
\end{figure}
\begin{deluxetable}{ccccc}
	\tablecolumns{5}
	\tablewidth{0pc}
	\tablecaption{Residual Atmospheric Dispersion}
	\tablehead{
		\colhead{} & \colhead{$\mathrm{d_x}$} & \colhead{$\mathrm{d_y}$}  & \colhead{PSF}\\
		\colhead{} & \colhead{} & \colhead{} & \colhead{spread}\\
		\cline{1-5}
		\colhead{} & \multicolumn{2}{c}{mas} & \colhead{\muas/nm}}
	\startdata
	Without Correction & -3.6 $\pm$ 8.3 & 41.6 $\pm$ 8.0 & 20.9 \\
	With Correction & 7.7 $\pm$ 6.5 & -5.3 $\pm$ 8.4 & 4.7 \\
	\enddata
	\footnotetext{$\mathrm{d_x}$ and $\mathrm{d_y}$ were calculated using the average and standard deviation from 1000~frame cube.}	\label{t:table}
\end{deluxetable}

The on-sky correction of residual atmospheric dispersion was achieved on the target Alpha Ari (spectral type K1, R-mag$=1.15$, H-mag$=-0.52$) on SCExAO's engineering night of October 30$^{th}$, 2015, using the same specifications. The results of the on-sky measurement and correction of residual atmospheric dispersion are presented in Fig.~\ref{f:result}. Figure~\ref{f:result} (a) shows the added speckles and the PSF core, with superimposed lines, showing that the radiation center lies away from the PSF core. Figure~\ref{f:result} (b) shows the added speckles after correcting for residual dispersion. Here the intersection point lies closer to the PSF core. After determining the residual dispersion and the appropriate angles of the ADC that would minimize it, the ADC was driven to a new position to make the radiation center coincide with the PSF core. The method outlined in Sec.~\ref{ssec:num1} was used to find the radiation center, and the results are presented in Tab.~\ref{t:table}. The values of $\mathrm{d_x}$ and $\mathrm{d_y}$ shown in the table represent the distance between the radiation center and the PSF core, in units of mas converted from pixel values (using the known scalling factor of 12.1~mas/pix in our NIR camera). The values of $\mathrm{d_x}$ and $\mathrm{d_y}$ are then converted in PSF elongation by the relationship between distance and dispersion given by Eq.~\ref{ratio}. Table~\ref{t:table} shows that in a single iteration, we have reduced the residual atmospheric dispersion from 20.9~\muas/nm to 4.7~\muas/nm, which corresponds to 4.2~mas in the y-H band and 1.4~mas in H-band alone, which is close to our science requirement of 1~mas in H-band. 

\section{Discussion}
The uncertainty in the values of $\mathrm{d_x}$ and $\mathrm{d_y}$ given in Tab.~\ref{t:table} were determined by calculating the standard deviation of those two values from a cube of 1000~images. This term consists of a combination of errors including the determination of the PSF core to sub-pixel accuracy as well as locating the radiation center which will be affected by the stability of the speckles due to both residual turbulence, possible blurring during an exposure, and the chromatic component to tip-tilt. A proper error budget is beyond the scope of this work but with careful consideration we believe the residual dispersion could be further optimized to the desired value of $<1$~mas.

One approach that could potentially improve the correction could be to exploit incoherent artificial speckles rather than the coherent versions used in this body of work. As explained in \citet{jovanovicsp15}, incoherent speckles can be created by rapidly modulating the phase of the artificial speckle grid by $\pi$ during a single exposure. In this way, speckles that do not interact with the underlying speckle halo can be created. Using these speckles, the stability of their brightness and centroid can be improved which would improve the determination of the radiation center. 

Finally, by using a method which uses the final science image to correct for residual atmospheric dispersion, it is possible to correct for dispersion resulting from optics internal to the instrument. This term is not taken into account in ADC models, and so direct measurement of this effect in the final focal plane is the only solution.

\section{Conclusion}\label{sec:summary}
In this work, we demonstrated the first on-sky measurement and correction of residual atmospheric dispersion using an adaptive speckle grid. This technique can be used by other AO systems, which do not employ high actuator DMs to create speckle grids, by using a diffractive grid in the pupil. This concept will be extended to work in conjunction with coronagraphs, where the location of the PSF behind the coronagraph can be found by cross-correlating the speckles in the grid. In this work, we have canceled one of the leading noise terms which prevents direct imaging of exoplanets. This work will be valuable in the field of ground-based high contrast imaging of habitable exoplanets in the era of ELTs, because it shows the path to correct atmospheric dispersion to the highest degree compared to traditional approaches. It can also be incorporated in atmospheric dispersion corrector design for ELTs \citep{bahrami}. 

The authors acknowledge support from the JSPS (Grant-in-Aid for Research 23340051 and 26220704).\\
\textit{Facility: Subaru Telescope.}

\end{document}